\documentclass[journal]{IEEEtran}

\usepackage[pdftex]{graphicx}
\usepackage{amsmath,bm}
\usepackage{amssymb}
\usepackage{soul,color}
\usepackage{listings}
\usepackage{leftidx}
\usepackage[T1]{fontenc}
\usepackage{hyperref}

\begin{document}
\title{Low-power switching of memristors exhibiting fractional-order dynamics}


\author{Nathan Astin and Yuriy~V.~Pershin,~\IEEEmembership{Senior~Member,~IEEE}
\thanks{N.~Astin and Y.~V.~Pershin are with the Department of Physics and Astronomy, University of South Carolina, Columbia, SC 29208 USA (e-mails: \mbox{nastin@email.sc.edu} and \mbox{pershin@physics.sc.edu}).}
\thanks{Manuscript received April ..., 2025; revised ....}}

\maketitle

\begin{abstract}
In this conference contribution, we present some initial results on switching memristive devices exhibiting fractional-order behavior using current pulses. In our model, it is assumed that the evolution of a state variable follows a fractional-order differential equation involving a Caputo-type derivative. A study of Joule losses demonstrates that the best switching strategy minimizing these losses depends on the fractional derivative's order and the power exponent in the equation of motion. It is found that when the order of the fractional derivative exceeds half of the power exponent, the best approach is to employ a wide pulse. Conversely, when this condition is not met, Joule losses are minimized by applying a zero current followed by a narrow current pulse of the highest allowable amplitude. These findings are explored further in the context of multi-pulse control. Our research lays the foundation for the advancement of the next generation of energy-efficient neuromorphic computing architectures that more closely mimic their biological counterparts.
\end{abstract}


\IEEEpeerreviewmaketitle

\section{Introduction} \label{sec:1}

Although it is typically assumed that a single time constant is sufficient to represent the behavior of a single neuron (like in the leaky integrate-and-fire model), power-law behaviors have been observed in a variety of biological systems~\cite{gilboa2005history,la2006multiple,tring2023power,vazquez2024fractional}. These kinds of behavior are termed scale-free because within a power-law distribution, a specific or typical scale is absent. Thus, electronic circuit components that exhibit a power law response are essential to precisely replicate biological neural systems in hardware.  Power-law behaviors are often associated with fractional derivatives~\cite{podlubny_fractional_1999}.

The motivation for replicating biological neural systems is because they possess several advantages over traditional AI systems (energy efficiency, adaptability, and advanced learning capabilities to name a few). Therefore, the ability to replicate biological neural systems has serious implications in the future of advanced computing and AI. With this in mind, this research aims to deepen the theoretical understanding of fractional-order electronics, focusing on a specific subset of resistors with memory (memristors) to comprehend the conditions under which they operate energy efficiently.

Traditionally, resistors with memory (also called memristors, ReRAM cells, memristive systems, etc.) are described using the memristive system equations introduced by Chua and Kang~\cite{chua76a}. In particular, the current-controlled memristive devices are defined by
\begin{eqnarray}
V_M(t)&=&R_M\left(\boldsymbol{x},I \right)I(t), \label{eq:1}\\
\dot{\boldsymbol{x}}&=&\boldsymbol{f}\left(\boldsymbol{x},I\right). \label{eq:2}
\end{eqnarray}
In the expressions above, $V_M$ and $I$ represent the voltage across and the current passing through the device, respectively, $R_M\left(\boldsymbol{x}, I\right)$ denotes the memristance (stands for memory resistance), which depends on the vector $\boldsymbol{x}$ consisting of $n$ internal state variables, while $\boldsymbol{f}\left(\boldsymbol{x}, I\right)$ is the vector state evolution function. Similarly, one could define voltage-controlled memristive devices as in \cite{chua76a}. According to Eq.~(\ref{eq:2}), in standard models, the evolution of internal state variables is governed by first-order differential equations.


In principle, a fractional derivative can be introduced in Eq.~(\ref{eq:1}) and/or Eq.~(\ref{eq:2}) in several ways.
In this work, we consider a modified version of Eq.~(\ref{eq:2}) with a fractional derivative on its left-hand side.
In this case, the model describes a purely resistive device exhibiting fractional-order dynamics. Alternative approaches include substituting the current $I(t)$ in Eq.~(\ref{eq:1}) with a fractional-order derivative of charge over time. This will result in a fractional order response that combines capacitive and resistive components (if the derivative order lies between zero and one). Devices with fractional-order response and fractional-order dynamics represent another interesting avenue to explore. Although fractional-order memristors have been studied in the past (see, e.g., Refs.~\cite{petras10a,fouda15a,Rajagopal17a,sanchez2018fractional,shatnawi23a,shi2018}), to the best of our knowledge, the problem of minimization of Joule losses has been considered only in the context of integer-order memristors~\cite{Slipko25a}.

This conference paper is organized as follows. Sec.~\ref{sec:2} presents the Caputo derivative, the Riemann-Liouville integral, and our memristor model. Joule losses are examined in Sec.~\ref{sec:3}. Our findings are discussed in Sec.~\ref{sec:4}. This section concludes with a summary and potential future research directions.

\section{Model of current-controlled memristors with fractional dynamics}\label{sec:2}


\begin{figure*}[tb]
(a)\includegraphics[width=0.44\textwidth]{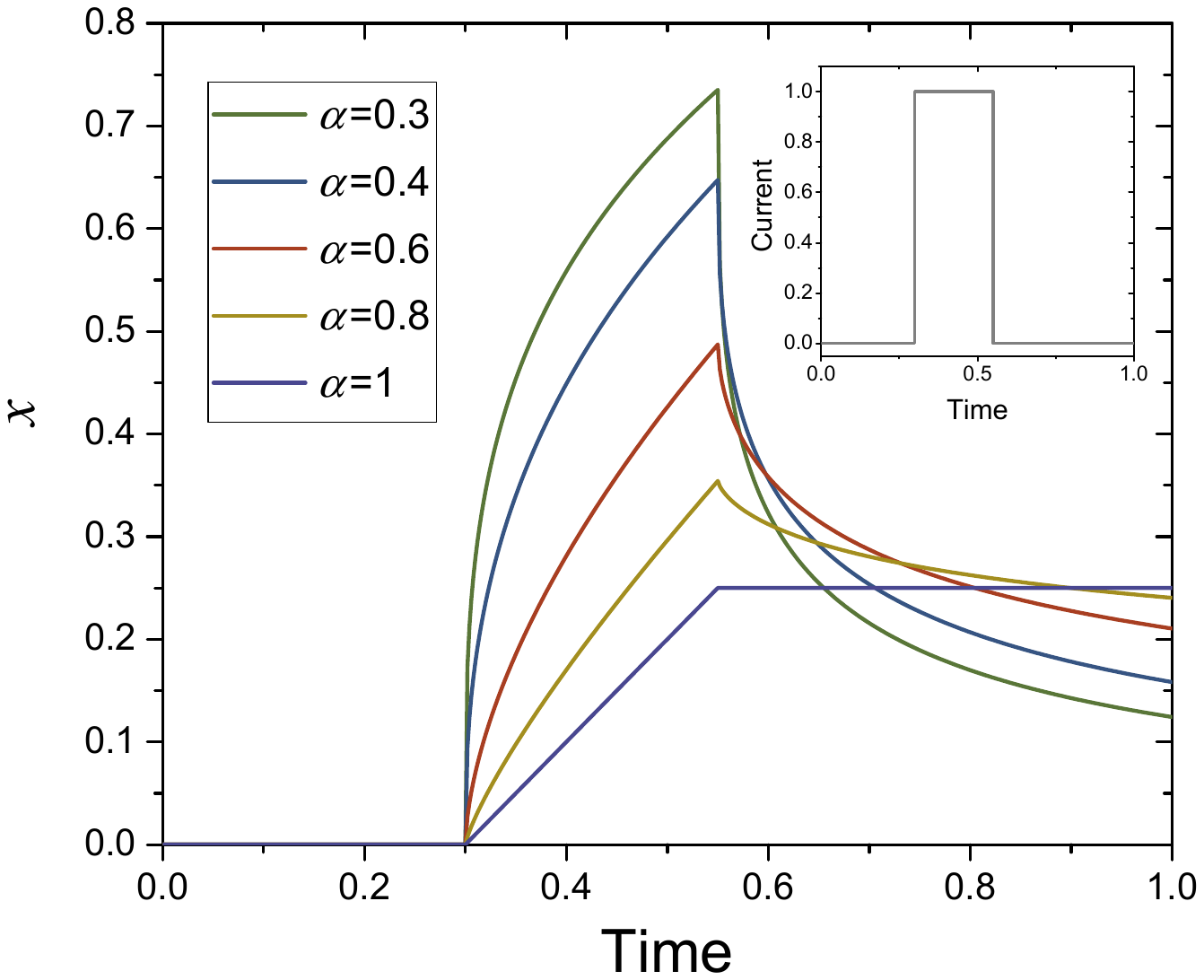} \;\;
(b)\includegraphics[width=0.44\textwidth]{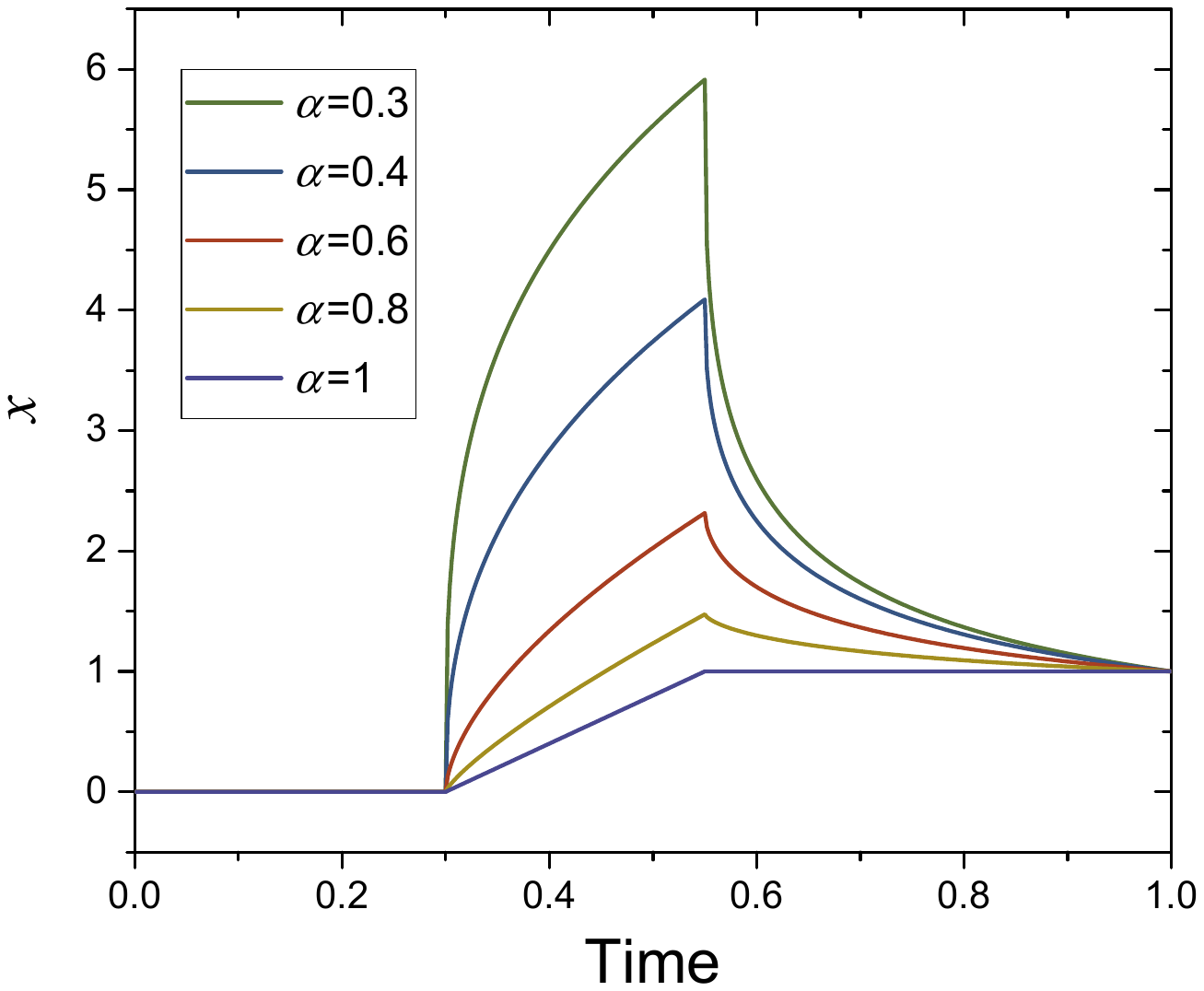}
\caption{
The response of the internal state variable to (a) same amplitude pulse (for all values of $\alpha$) and (b)
variable amplitude pulse. These graphs were obtained using $A=1$, $B=5$, $x_0=0$, $t_{st}=0.3$, $t_e=0.55$, $\beta=1$, and $I_1=1$ (in (a)). To obtain (b), we imposed the constraint $x_1(t=1\equiv t_1)=1$ and used Eq.~(\ref{eq:8}) to calculate the pulse amplitude for each value of $\alpha$. The pulse waveform used in (a) is presented in its inset.
\label{fig:1}}
\centering
\end{figure*}

Although there exist many definitions of fractional derivative~\cite{teodoro2019review}, here we consider the Caputo fractional derivative~\cite{caputo1967linear}. The advantage of the Caputo derivative (compared, for instance, to the Riemann-Liouville fractional derivative) is that it leads to initial conditions expressed through integer-order derivatives. As a result,  the fractional version of the fundamental theorem of calculus for Caputo derivatives is of the same form as the fundamental theorem of calculus for integer order \cite{grigoletto2013}.
The Caputo derivative is defined by the expression
        \begin{equation}
        \leftidx{^C_a}{D_{t}^\alpha}{f(t)}=\frac{1}{\Gamma(n-\alpha)}\int_{a}^{t}\frac{f^{(n)}(\tau)}{(t-\tau)^{\alpha+1-n}}\,\textnormal{d}\tau\; , \label{eq:3}
        \end{equation}
where $\Gamma(...)$ denotes the gamma function, $\alpha$ is the order of the derivative, and $n$ is defined by $n-1<\alpha<n \in \mathbf N$. For future use, we also provide the left Riemann-Liouville integral:
        \begin{equation}
           \leftidx{_a}{J_{t}^\alpha} {f(t)}=\frac{1}{\Gamma(\alpha)}\int_{a}^{t} (t-\tau)^{\alpha-1}f(\tau)\,\textnormal{d}\tau . \label{eq:4}
        \end{equation}
These definitions can be found, for example, in Ref.~\cite{podlubny_fractional_1999} and other literature. To simplify the presentation, we assume that all the equations are expressed in dimensionless variables.

In our model, the memristor satisfies the generalized Ohm's law (Eq.~(\ref{eq:1})), and its memristance is a linear function of the internal state variable, $x$, according to
\begin{equation}
    R_M(x)=A+Bx, \label{eq:5}
\end{equation}
where $A$ and $B$ are positive constants. The only constraint on $x$ in our model is $x\geq 0$, which guarantees that the memristance is positive. Eq.~(\ref{eq:2}), however, is replaced with a fractional-order differential equation
\begin{equation}
    \leftidx{^C_0}{D_{t}^\alpha}{x(t)}=\kappa \cdot \textnormal{sign} (I) \left|I\right|^\beta, \label{eq:6}
\end{equation}
where $\kappa$ and $\beta$ are positive constants. We emphasize that  Eq.~(\ref{eq:6}) utilizes a particular form for the state evolution function, which, in principle, may have a different functional form. However, in the case of $\alpha=\beta=1$, this model corresponds to the TiO$_2$ memristor model introduced in \cite{missingmemristor}. In \cite{missingmemristor}, the switching is explained by the drift of oxygen  vacancies across the device that changes the width of
an oxygen deficient TiO$_{2-x}$ layer. The doped (oxygen vacancy rich) and undoped layer effectively act as a pair of resistors in series, and the width of the doped layer in relation to the device width translates to the device's memristance. Therefore, in accordance with this model, the left side of Eq.~\eqref{eq:6} at $\alpha=1$ corresponds to the drift velocity of the doped/undoped boundary interface. This model uses only integer-order dynamics; however, the inclusion of fractional-order dynamics incorporates memory effects.

\begin{figure*}[t]
(a) \includegraphics[width=0.44\textwidth]{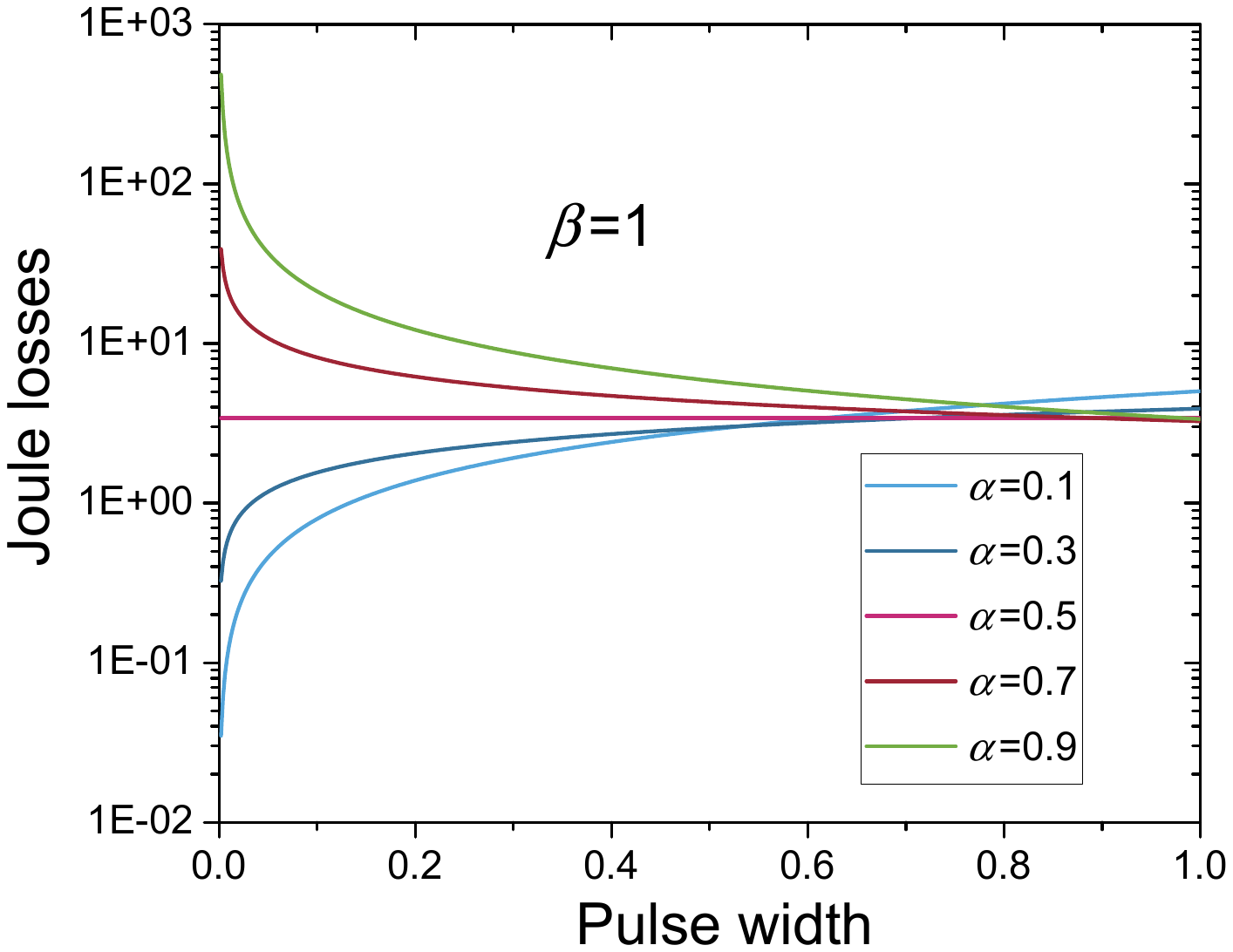}  \; (b) \includegraphics[width=0.44\textwidth]{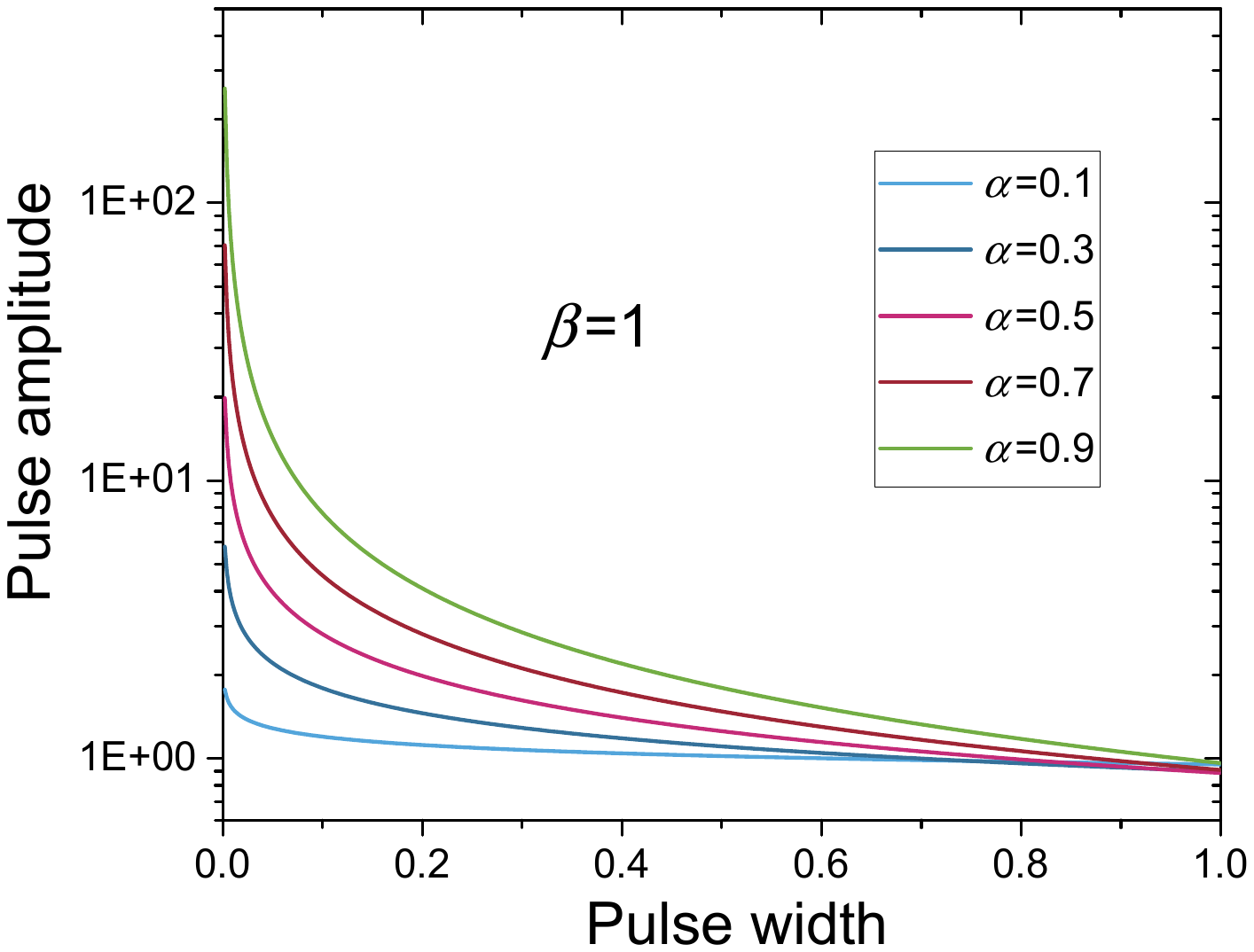}
\\
(c) \includegraphics[width=0.44\textwidth]{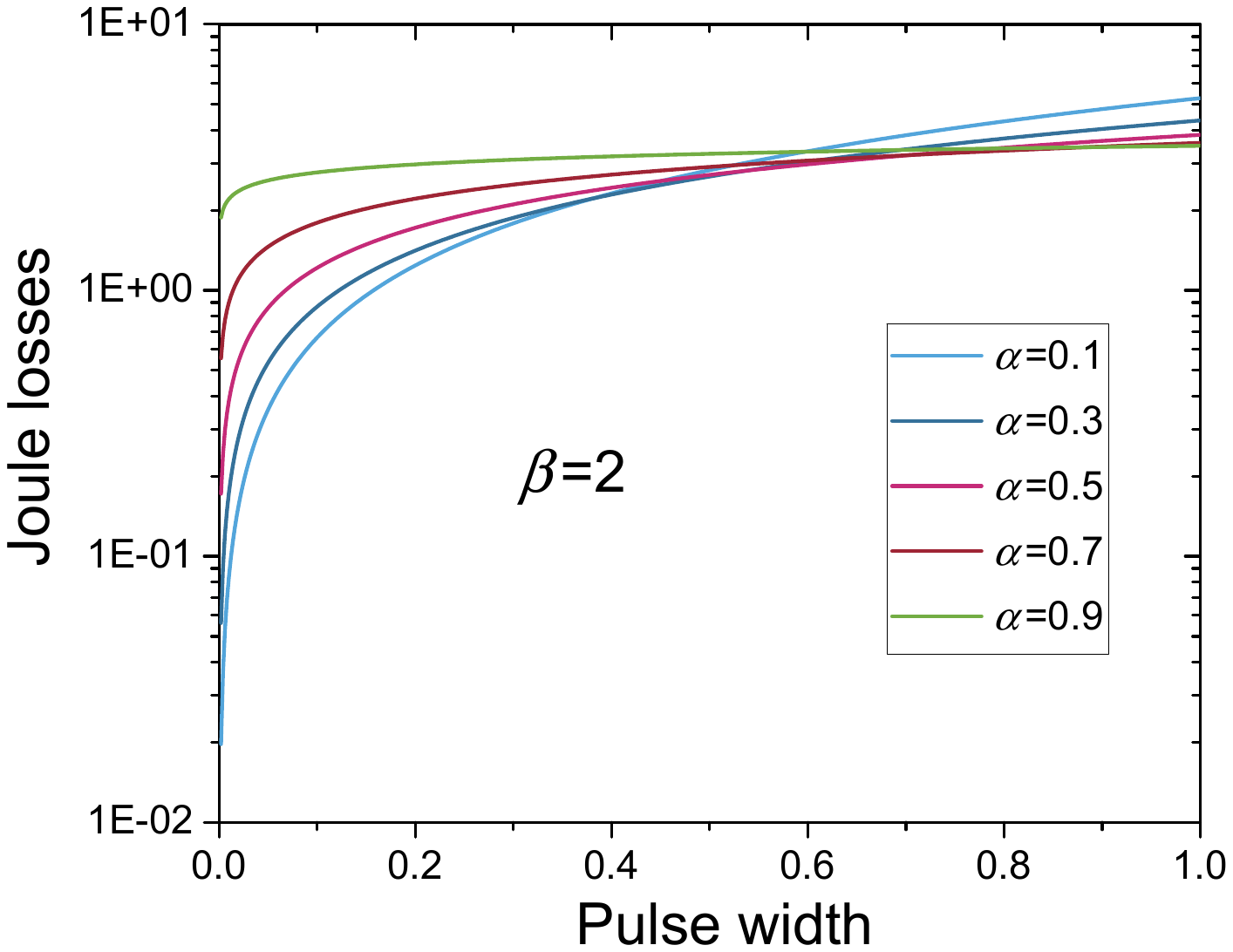}  \; (d) \includegraphics[width=0.44\textwidth]{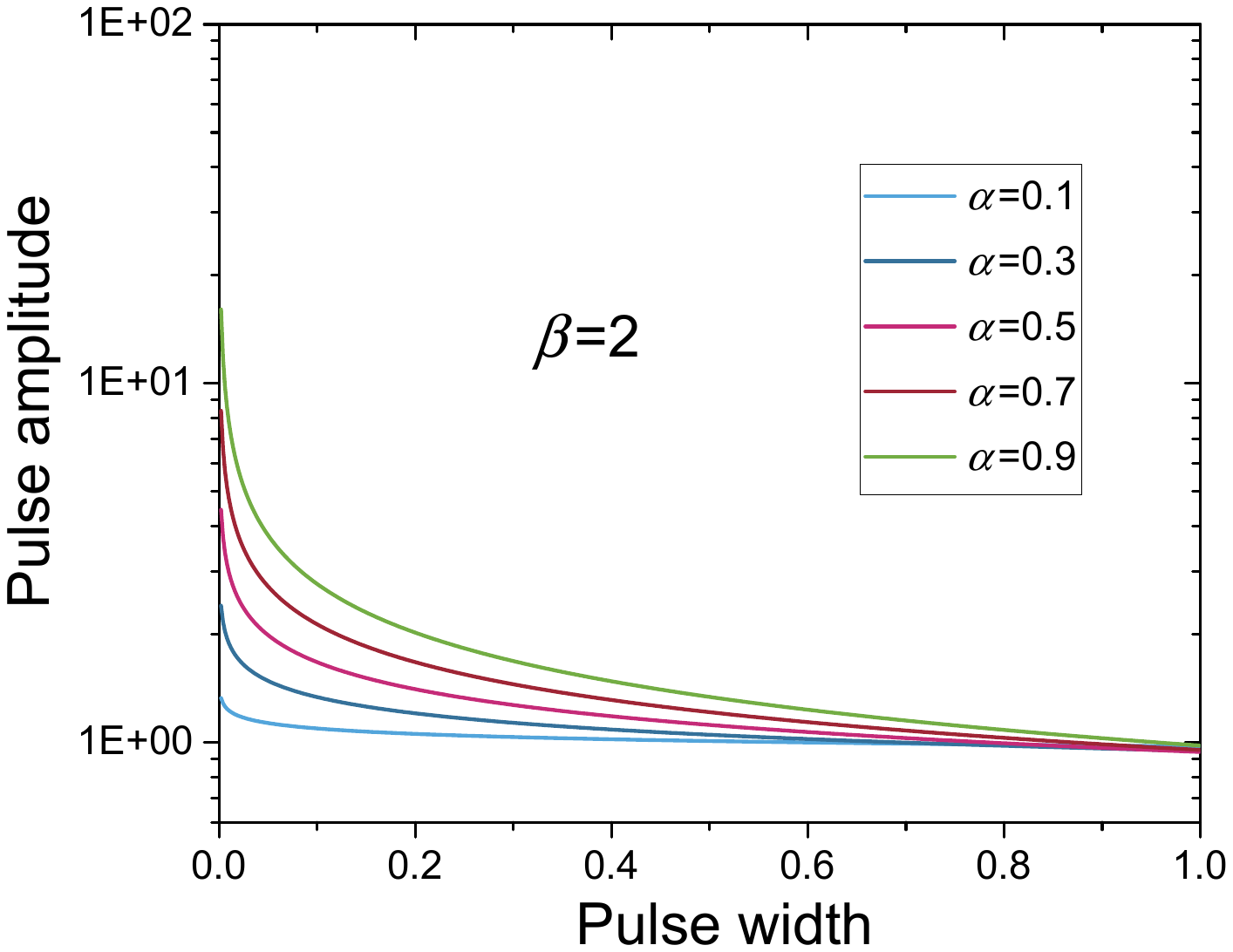}
\caption{Switching by a single pulse. (a), (c) Joule losses and (b), (d) corresponding pulse amplitudes as functions of the pulse width $T=t_e-t_{st}$ calculated for selected values of $\alpha$ and $\beta$ using $t_e=t_1=1$.  All other parameter values are the same as in Fig.~\ref{fig:1}.
\label{fig:2}}
\centering
\vspace{-0.2 cm}
\end{figure*}

Assuming that the memristor is subjected to a constant amplitude current pulse defined by the start time, $t_{st}\geq 0$, end time, $t_{e}>0$, and amplitude, $I_1>0$, Eq.~(\ref{eq:6}) can be integrated by the application of the left Riemann-Liouville integral, Eq.~(\ref{eq:4}). This leads to
\begin{eqnarray}
&& x(t)-x_0 =\nonumber \\
&& \begin{cases}
 0&\text{for $t\leq t_{st}$} \; ,\\
\frac{\kappa I_1^\beta}{\Gamma(\alpha+1)}(t-t_{st})^\alpha &\text{for $t_{st}\leq t\leq t_{e}$} ,\\
\frac{\kappa I_1^\beta}{\Gamma(\alpha+1)}\left[(t-t_{st})^\alpha-(t-t_e)^\alpha\right] &\text{for $t_{e} \leq t$} \; .
\end{cases} \qquad
\label{eq:7}
\end{eqnarray}
The above solution is valid for $0<\alpha\leq 1$. Fig.~\ref{fig:1}(a) exemplifies the solution (\ref{eq:7}) for several selected values of $\alpha$. The shape of the pulse used in the calculations is shown in the inset of Fig.~\ref{fig:1}(a). An interesting feature of the solution (\ref{eq:7}) is the relaxation of $x$ to its initial value, $x_0$, at $t>t_e$ for all $0<\alpha<1$. There is no relaxation at $\alpha=1$.


\begin{figure*}[t]
(a) \includegraphics[width=0.44\textwidth]{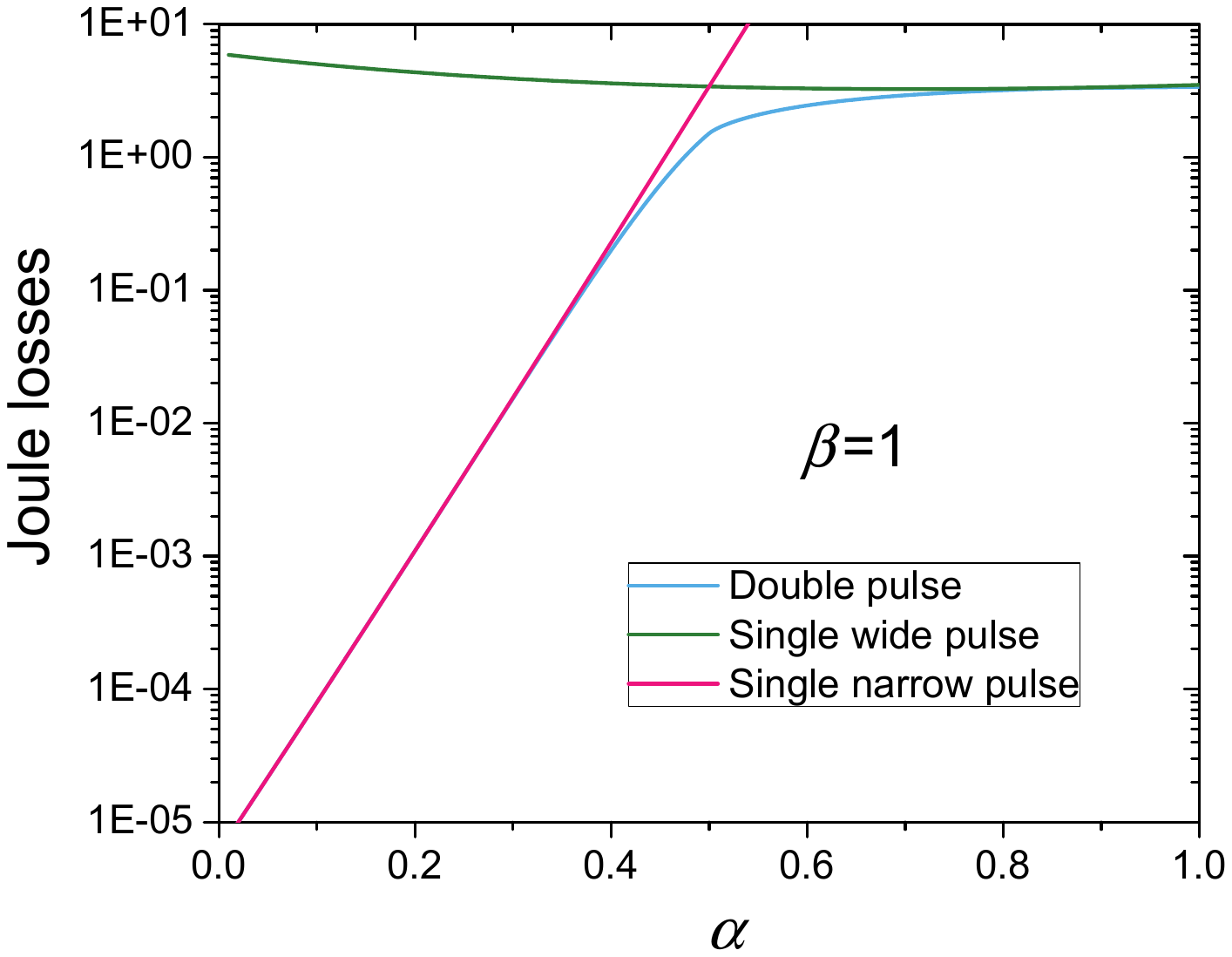}  \; (b) \includegraphics[width=0.49\textwidth]{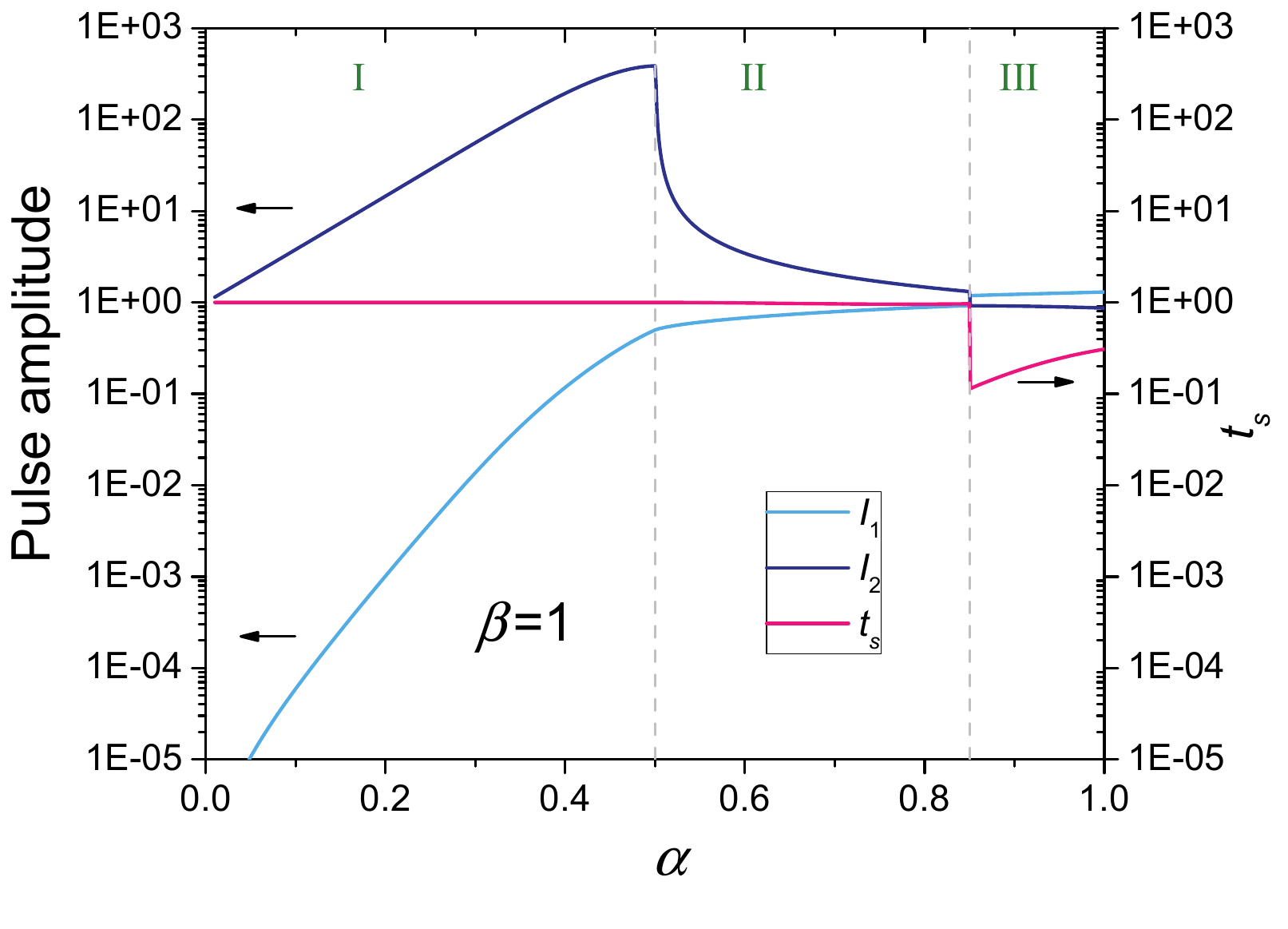}
\\
(c) \includegraphics[width=0.44\textwidth]{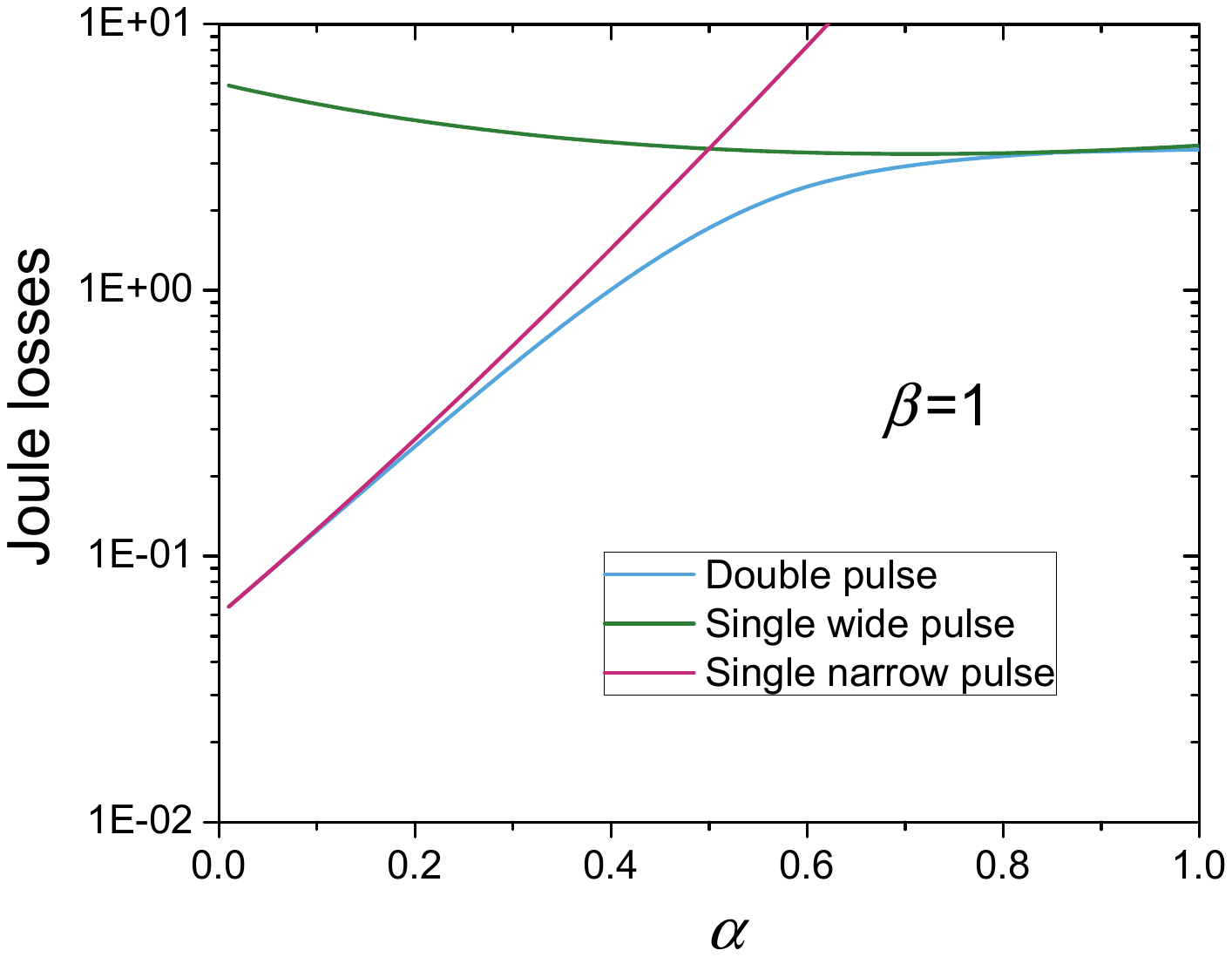}  \; (d) \includegraphics[width=0.49\textwidth]{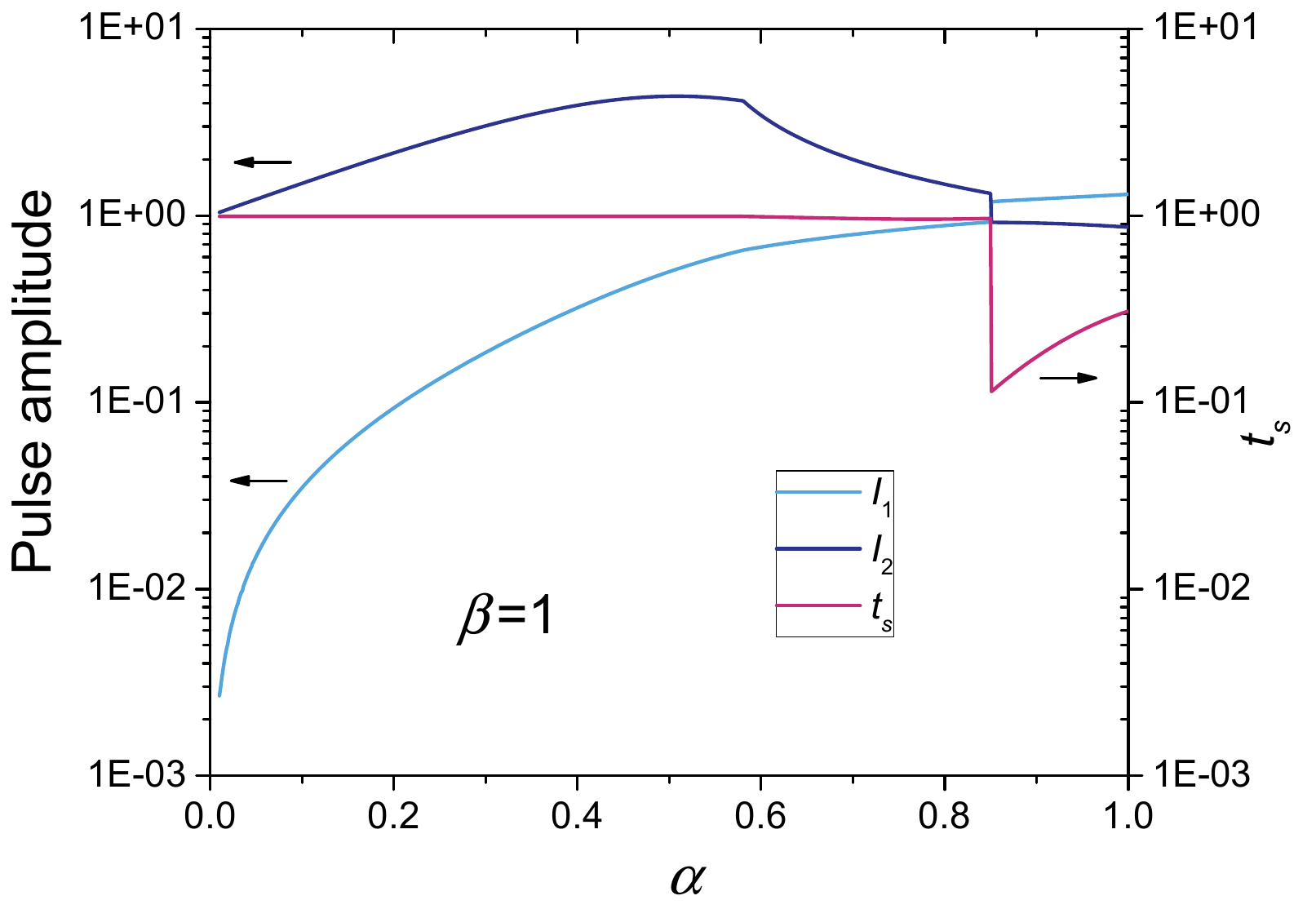}
\caption{Switching by two pulses. (a), (c) Optimal Joule losses and (b), (d) corresponding pulse parameters as functions of $\alpha$ at $\beta=1$.  All other parameter values are the same as in Fig.~\ref{fig:1}. The pulse sequence is described in the text. The dashed vertical lines in (b) correspond to $\alpha=0.5$ and $\alpha=0.85$. (a) and (b) were obtained assuming the minimal pulse width $\textnormal{min}\{ T_1,T_2\}=10^{-6}$ while for (c), (d) we used $\textnormal{min}\{ T_1,T_2\}=10^{-2}$. For comparison purposes,  in (a) and (c) some single pulse curves are shown. These curves are defined in the text.
\label{fig:3}}
\centering
\end{figure*}

\section{Results}\label{sec:3}

\subsection{Single pulse}

In this subsection, we use a single square pulse of current to switch a memristor from $x(0)=x_0$ to $x(t_1)=x_1$, where $t_1$ is the final moment of time. To find the required pulse amplitude, $I_1$, assuming that the start and end times for the pulse are given, we consider the last line on the right-hand side of Eq.~(\ref{eq:7}) at $t=t_1$ and solve Eq.~(\ref{eq:7}) with respect to $I_1$. The result is
\begin{equation}
    I_1=\left[\frac{1}{\kappa} \cdot \frac{\Gamma(\alpha+1)(x_1-x_0)}{(t_1-t_{st})^\alpha-(t_1-t_e)^\alpha} \right]^{\frac{1}{\beta}} \; . \label{eq:8}
\end{equation}

Fig.~\ref{fig:1}(b) illustrates trajectories obtained using Eqs.~(\ref{eq:7}) and (\ref{eq:8}) (note the same final state). An interesting aspect is that these trajectories remain unchanged regardless of the parameter $\beta$. This becomes evident when substituting Eq.~(\ref{eq:8}) into Eq.~(\ref{eq:7}). For instance, the bottom line on the right-hand side of Eq.~(\ref{eq:7}) can be rewritten as
\begin{equation}
\left(x_1-x_0\right) \frac{(t-t_{st})^\alpha-(t-t_e)^\alpha}{(t_1-t_{st})^\alpha-(t_1-t_e)^\alpha}.\label{eq:9}
\end{equation}
Moreover, Eq.~(\ref{eq:9}) shows that at $\alpha=1$, $x(t)$ remains constant at $x_1$ when $t>t_e$. This is what happens in the traditional model described by Eq.~(\ref{eq:2}) as it should be.

Next, consider the Joule losses. Using the linear memristance model, Eq.~(\ref{eq:5}), one can easily derive
\begin{eqnarray}
   \nonumber  Q&=&\int\limits_{t_{st}}^{t_e}I_1^2\left(A+Bx_0+B\frac{\kappa I_1^\beta}{\Gamma(\alpha+1)}(t-t_{st})^\alpha \right)\textnormal{d}t=\\
    && I_1^2\left[\frac{A+Bx_0}{(t_e-t_{st})^\alpha}+B\frac{\kappa I_1^\beta}{\Gamma(\alpha+2)}\right](t_e-t_{st})^{\alpha+1}
    \; \label{eq:10}
\end{eqnarray}
with $I_1$ defined by Eq.~(\ref{eq:8}).

Fig.~\ref{fig:2} shows the Joule losses (calculated using Eq.~(\ref{eq:10})) and pulse amplitudes (calculated using Eq.~(\ref{eq:8})) for a set of selected values of the parameters $\alpha$ and $\beta$. Given that fractional-order device dynamics include relaxation (illustrated in Fig.~\ref{fig:1}), we focus exclusively on the pulses applied late in the interval of interest, assuming that they offer the greatest energy efficiency.
Fig.~\ref{fig:2}(a) shows that at $\beta=1$, Joule losses are independent of the pulse width at $\alpha=0.5$, decrease with the pulse width when $\alpha>0.5$, and increase with the pulse width when $\alpha<0.5$. According to Fig.~\ref{fig:3}(c), at $\beta=2$ in all cases, Joule losses increase with pulse width.
 Fig.~\ref{fig:2}(a) and (c) suggests the existence of an optimal strategy for controlling memristors exhibiting fractional-order dynamics.

To better understand the behaviors in Fig.~\ref{fig:2}, we note that when $t_e=t_1$, Eq.~(\ref{eq:10}) becomes
\begin{eqnarray}
    \nonumber Q&=&\left(\frac{\Gamma(\alpha+1)(x_1-x_0)}{\kappa} \right)^{\frac{2}{\beta}} \cdot  \\
    && \left(A+B x_0 + \frac{B(x_1-x_0)}{\alpha+1}\right) \left( t_1-t_{st} \right)^{1-\frac{2\alpha}{\beta}}.\;\; \label{eq:11}
\end{eqnarray}
A key observation is that
\begin{equation}
 \lim\limits_{(t_1-t_{st})\to 0^+} Q=0 \;\;\; \textnormal{when} \;\;\; \alpha<\frac{\beta}{ 2}\;. \label{eq:12}
\end{equation}

\subsection{Two pulses}

In this subsection, we explore the optimal switching control by using two back-to-back current pulses, which we also call a double pulse. The pulses are assumed to span the entire interval from $t=0$ to $t_1$ so that $t_{st,1}=0$, $t_{e,1}=t_{st,2}$, $t_{e,2}=t_1$, where the indices 1 and 2 refer to the first and second pulse. Our goal is to find the optimal values of three ``free'' parameters -- the pulse amplitudes ($I_1$ and $I_2$) and the switching time from the first to the second pulse ($t_s\equiv t_{e,1}=t_{st,2}$). Due to the complexity of analytical expressions, the optimal pulse parameters were found numerically using Mathematica 14.2.

First, we find the trajectory of the internal state, $x(t)$, by solving Eq.~(\ref{eq:6}). The result is
\begin{eqnarray}
&&x(t)-x_0 = \nonumber \\
&&\begin{cases}
\frac{\kappa I_1^\beta}{\Gamma(\alpha+1)}t^\alpha &\text{for $t\leq t_s$} ,\\
\frac{\kappa I_1^\beta(t^\alpha-(t-t_s)^\alpha)+\kappa I_2^\beta(t-t_s)^\alpha)}{\Gamma(\alpha+1)} & \text{for $t_s\leq t\leq t_1$}
\end{cases}. \qquad
\label{eq:77}
\end{eqnarray}
In order to ensure that the same final state is reached after applying the second current pulse, \(I_2\) must take the form of
\begin{eqnarray}
   \nonumber I_2(I_1,t_s)&=&\left[\frac{(t_1-t_s)^{-\alpha}}{\kappa}\Gamma(1+\alpha) \right]^{\frac{1}{\beta}}\cdot\\&&\cdot\left( x_1-x_0+\frac{\kappa I_1^\beta\left[ (t_1-t_s)^\alpha-t_1^\alpha\right]}{\Gamma(1+\alpha)}\right)^\frac{1}{\beta}. \;\;\; \label{eq:14}
\end{eqnarray}
Eq.~(\ref{eq:14}) simplifies the problem by cutting down the number of unknown parameters to just two. Consequently, in the subsequent analysis, we focus on optimizing $Q$ with respect to $I_1$ and $t_s$.

Taking the integral of the power function from 0 to $t_1$ (assuming the initial and final states are 0 and 1 respectively), the expression for $Q(I_1,t_s)$ takes the form of
\begin{eqnarray}
   \nonumber Q(I_1,t_s)&=&\frac{(t_1-t_s)^{\frac{-2\alpha}{\beta}}}{\Gamma(2+\alpha)}[ BI_1^\beta \kappa(t_1^\alpha-t_s^\alpha)t_s+\\
   \nonumber  &&+(t_1-t_s)\left(A(1+\alpha)+B\right)\Gamma(1+\alpha)]\cdot\\
   \nonumber &&\cdot\left(I_1^\beta\left( (t_1-t_s)^\alpha-t_1^\alpha\right)+\frac{\Gamma(1+\alpha)}{\kappa} \right)^{\frac{2}{\beta}}+\\&&+I_1^2t_s\left( A+B\frac{\kappa I_1^\beta t_s^\alpha}{\Gamma(2+\alpha)}\right)\label{eq:15}.\;\;
\end{eqnarray}

In what follows, $Q(I_1,t_s)$ given by Eq.~(\ref{eq:15}) is numerically minimized to obtain the optimal values of $I_1$ and $t_s$.

Fig.~\ref{fig:3}(a) and (c) presents an example of our results showing the optimized values of $Q$ depending on the derivative order $\alpha$. The optimization process was carried out with a restriction on the minimum pulse width, as detailed in the caption of Fig.~\ref{fig:3}. Fig.~\ref{fig:3} shows the optimal control parameters in panels (b) and (d), respectively. The single pulse curves in Fig.~\ref{fig:3}(a) and (c) were calculated using Eq.~(\ref{eq:11}) with $t_s$ equal to $0$ and minimal pulse width.

When the minimal pulse width is very small, as in Fig.~\ref{fig:3}(a) and (b), one can clearly distinguish three well-defined control regimes. The first regime (I) is characterized by the minimal width of $T_2$ and a large ratio of $I_2/I_1$. The end of regime I is characterized by a drop of $I_2$ precisely at $\alpha=0.5$. Regime III starts with a drop of $t_s$ at $\alpha\approx 0.85$. In this regime, the values of $I_1$ and $I_2$ are relatively close to each other, however, their order is changed (compared to regimes I and II). The second regime lies between regimes I and III.

Fig.~\ref{fig:3}(c) and (d) demonstrates how a larger value of the minimal pulse width modifies the curves in Fig.~\ref{fig:3}(a) and (b). Fig.~\ref{fig:4} shows the evolution of $x$ for different values of $\alpha$ and is discussed in Sec.~\ref{sec:4} below.

\begin{figure}[tb]
\centering
\includegraphics[width=0.44\textwidth]{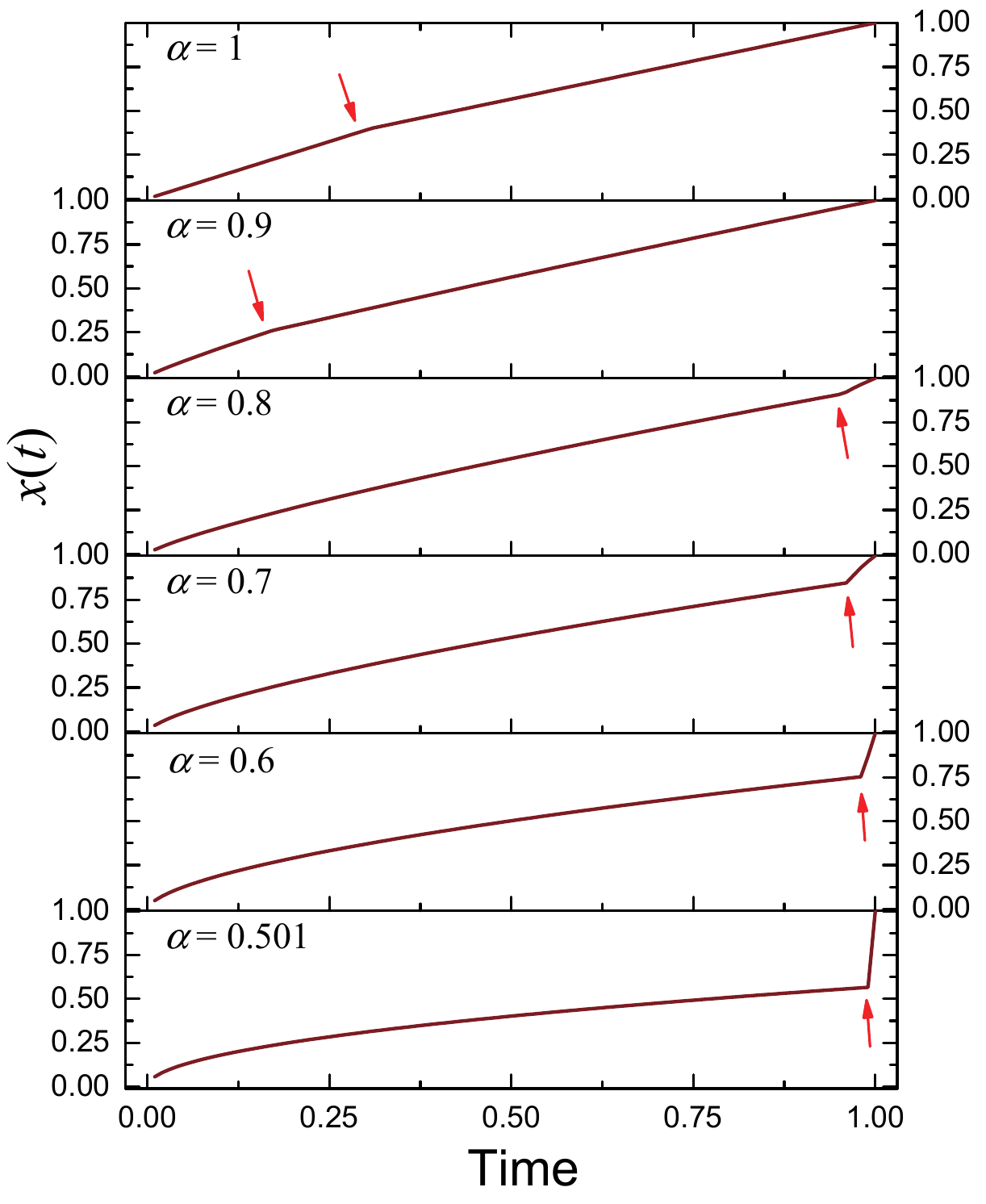}
\caption[width=\linewidth]{The response of the internal state variable for \(\beta=1\) of various \(\alpha \) using the optimal choice of parameters, \(I_1\) and \(t_s\), for a minimum Q. The arrows represent the moments of the transition from the first to second pulse.
\label{fig:4}}
\centering
\end{figure}

\begin{figure}[tb]
\centering
 \includegraphics[width=0.44
 \textwidth]{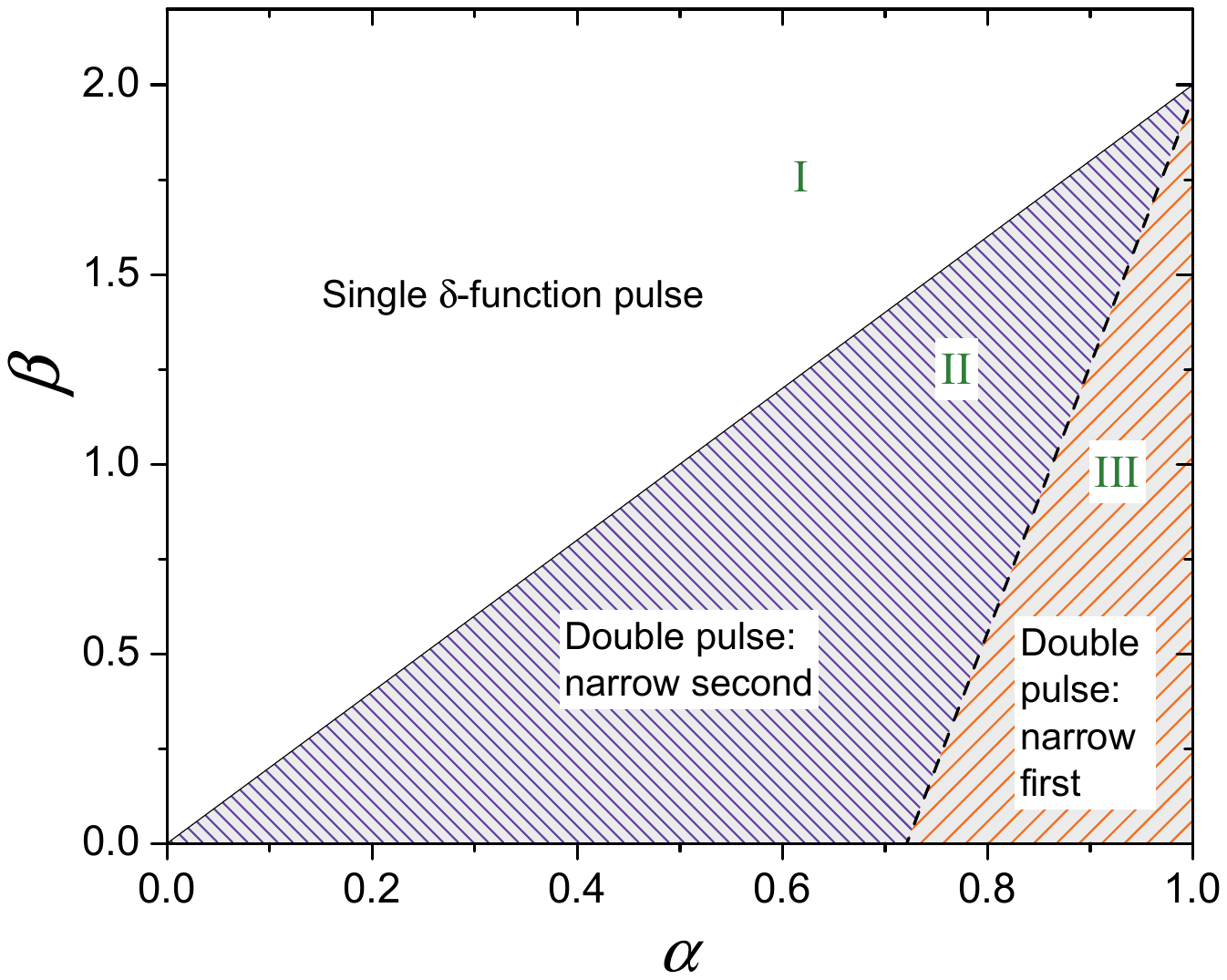}
\caption[width=\linewidth]{Phase diagram for optimal choice of pulses. The boundary lines are \(\beta=2\alpha\) as determined by the single pulse case and 
\(\beta\approx7.05\alpha-5.08\) found numerically with the double pulse expression for $Q$, Eq.~\eqref{eq:15}, using the same parameters as used in Figs.~\ref{fig:1}-\ref{fig:3}. In the case of double pulses, the boundary is a separation between when a narrow second pulse versus a narrow first pulse is more optimal.
\label{fig:5}}
\centering
\end{figure}

While Fig.~\ref{fig:3} and Fig.~\ref{fig:4} focus on the case of $\beta=1$, we have also performed extensive numerical simulations to understand the best approaches of control in the $\alpha-\beta$ phase space. Fig.~\ref{fig:5} shows the result of these simulations.
According to Fig.~\ref{fig:5}, when $\beta<2$ there exist three regimes of optimal control (previously identified in Fig.~\ref{fig:3}(b)). The boundaries between these regimes show quite a linear dependence on the derivative order $\alpha$. When $\beta>2$, the only optimal approach involves the use of a narrow pulse at the maximum permissible amplitude.

\section{Discussion and conclusion} \label{sec:4}

In this conference paper, we have explored low-power pulse control of memristive devices with fractional-order dynamics based on a particular device model.
We have observed that the energy cost to switch the device state depends on the driving protocol, specifically, the amplitude and time parameters of a pulse or pulses.

In the case of a single square pulse, we have found two optimal control regimes. When the order of the fractional derivative exceeds half of the power exponent, the best approach is to employ a wide pulse. Conversely, when this condition is not met, Joule losses are minimized by applying a zero current followed by a narrow current pulse of the highest allowable amplitude. This result agrees with previous work~\cite{Slipko25a} on integer-order memristors where the same approach was anticipated for the $\alpha=1$ case.

In the case of a double pulse control, we have determined three optimal control regimes. The first regime where the order of the fractional derivative is less than half of the power exponent is the same as for the single pulse due to the limit of zero pulse width seen in Eq. \eqref{eq:12}. When this first condition is not met, we see a deviation from the single pulse result.  Each of these regimes contains one narrow and wide pulse, and in both regimes the majority of Joule losses comes from the application of the wide pulse. The second regime corresponds to a wide first pulse followed by a narrow second pulse. Conversely, the third regime corresponds to a narrow first pulse followed by a wide second pulse. It can be seen in Fig.~\ref{fig:5} that the final regime begins at \(\alpha \approx0.72\), and the necessary condition to belong to this regime is that, approximately, $\beta<7\alpha-5$.
When this condition is not met, Joule losses are minimized by applying a wide first pulse followed by a narrow second.

The differentiation between regimes II and III can be understood qualitatively through the interplay of two mechanisms. The first mechanism is the relaxation in fractional cases (refer to Fig.~\ref{fig:1}), which is more significant at lower $\alpha$ values. The other mechanism is the tendency to distribute power roughly evenly over time~\cite{Slipko25a}, aiming to reduce Joule losses~\footnote{In the context of ideal memristors, the ideal power remains constant, as articulated in Theorems 1 and 2 in~\cite{Slipko25a}.}. Regime II is characterized by smaller $\alpha$ values where the relaxation effect dominates. Hence, it is advantageous to apply the higher amplitude pulse second. In contrast, for larger $\alpha$ values (regime III), the significance of relaxation diminishes. Consequently, it's more effective to apply a stronger pulse initially while the resistance is lower. Refer to Fig.~\ref{fig:3}(b) and (d) for sample pulse amplitudes. Additionally, the relative magnitudes of pulses can be inferred from the gradient of $x(t)$ shown in Fig.~\ref{fig:4}.

An important result can be seen in Fig.~\ref{fig:3}(a) and (c) that when only $\alpha$ is changed, the Joule losses decrease as the order of the fractional derivative decreases. As a result, switching memristive devices described by fractional dynamics in Eq.~\eqref{eq:6}  from an initial state to a final state seems to be more energy efficient than switching systems described with integer-order dynamics (at the same numerical values of parameters). It follows from this and the presence of power-law behaviors in biological systems that fractional memristive devices may be better suited to replicating the energy efficiency of biological neural systems.

As stated above, at $\alpha=\beta=1$ our model corresponds to the linear dopant drift model introduced in \cite{missingmemristor}. Unfortunately, this linear (ideal) memristor model faces serious limitations~\cite{Ventra_2013}. For example, even small voltages can lead to large electric fields in nanoscale devices, which leads to non-linearities in ionic transport \cite{missingmemristor}. Therefore, non-linear models, such as the one presented in \cite{6353604}, provide a more accurate description of physical memristive devices.

Due to the potential benefits replicating the brain's energy efficiency would entail, this topic warrants further research. There are several interesting future directions one could pursue. For one, this pulse control approach can be easily extended to other memristance models using the same fractional dynamics. Alternatively, an arbitrary number of current pulses could be considered with the goal of finding an optimal number of pulses for a particular current-controlled device. In addition, a large number of pulses could be used to approximate a continuous solution providing a lower bound on the energy efficiency for switching.  Furthermore, it may be interesting to explore how the phase space \(\alpha-\beta\) changes when enforcing the condition of a minimum pulse width or threshold current. It would also be interesting to better understand the reason for the sharp transition between regimes II and III seen in Fig.~\ref{fig:3}(b) and (d). Further, this work can only be compared to physical devices in a limited scope, it is necessary to then consider models with nonlinear fractional dynamics. In addition to current-controlled devices, one could also explore the low-power switching of voltage-controlled devices with fractional-order dynamics.

\section*{Acknowledgement}
The authors thank Fidel Santamaria for an interesting suggestion. This research was supported by the NSF grant EFRI-2318139.


\end{document}